\documentclass[a4paper,fleqn]{cas-sc}

\usepackage[authoryear,longnamesfirst]{natbib}
\usepackage{subcaption}
\def\tsc#1{\csdef{#1}{\textsc{\lowercase{#1}}\xspace}}
\tsc{WGM}
\tsc{QE}

\begin{document}
\let\WriteBookmarks\relax
\def\floatpagepagefraction{1}
\def\textpagefraction{.001}

\shorttitle{}    

\shortauthors{}  

\title{Title Page}
\title [mode = title]{JiRAIYA, A Decentralized Hierarchical Federated Learning Framework}

\author[1]{Venkata Raghava Kurada}
\credit{Methodology, Software Developemnt}
\ead{venkataraghavakurada@sssihl.edu.in}
\cormark[1]
\author{Pallav Kumar Baruah}

\ead{venkataraghavakurada@sssihl.edu.in}
\ead{pkbaruah@sssihl.edu.in}

\credit{Conceptualization of this study, Supervision}

\affiliation{organization={Dept of Mathematics and Computer Science, Sri Sathya Sai Institute of Higher Learning},
            addressline={Prashanthi Nilayam Campus, Vidhyagiri}, 
            city={Puttaparthi},
            postcode={515134}, 
            state={Andhra Pradesh},
            country={India}}

\begin{abstract}
Federated Learning(FL) is predominantly deployed in enterprise environments, where limited transparency and restricted auditability hinder broader adoption. Existing FL systems often suffer from opaque aggregation processes, making it unclear which model updates are accepted or discarded. Current mitigation strategies typically rely on external validators introducing additional computational and communication overhead.
In this paper, we propose a novel FL framework that leverages existing Web3 technologies to enhance transparency, trust and auditability throughout the training process. The framework adopts a hierarchical architecture in which delegated managers orchestrate the FL training process within their respective federations. To mitigate adversarial and poisoning attacks, a combination of novelty detection and consensus mechanisms were employed. Model updates are encoded and broadcasted to all managers, who independently evaluate their validity and those model updates that are approved by the consensus are incorporated into the global model. Additionally, a reputation score based backup mechanism is employed to ensure model generation.
Extensive experiments conducted under real world scenarios demonstrate the effectiveness, resilience of the proposed framework, highlighting its potential to enable transparent FL beyond traditional enterprise setting.
\end{abstract}


\begin{highlights}
\item  \textbf{Efficient and Privacy-Focused Federated Learning Implementation for Web3:}
An efficient federated learning framework is introduced for Web3 environments with a strong emphasis on minimizing  gas consumption. The design carefully limits smart contract interactions and reduces on-chain data footprint, enabling practical deployment under real world blockchain cost constraints. To ensure privacy, raw model updates are never exposed. Instead, all local updates are encoded and encrypted prior, preventing information leakage and protecting participant data throughout the training process.

\item \textbf{Scalable and Near-Instantaneous Malicious Update Mitigation via Novelty Detection and Consensus:}
A novel malicious update mitigation mechanism is presented that combines statistical novelty detection with a decentralized consensus-based validation process. This approach enables near-instantaneous identification of anomalous or adversarial updates while avoiding expensive iterative verification. The streamlined validation workflow is highly scalable, allowing secure and robust federated learning even in large, dynamic, and permissionless Web3 settings.
\end{highlights}

\maketitle

\maketitle

\section{Introduction}\label{sec1}
ML has become a transformative force across industries, fundamentally reshaping how data is analyzed and how intelligent systems are built. Advances in data collection, the widespread availability of large scale datasets and increased access to affordable and powerful computational resources have significantly lowered the barriers to developing complex ML models. As a result, ML applications now permeate domains ranging from healthcare and finance to smart cities and personalized services. While this diverse data offers new opportunities, it also exposes fundamental limitations of traditional centralized ML training approaches. In conventional settings, raw data is collected from multiple sources and aggregated at a centralized data center for training, raising serious concerns related to data privacy, ownership and regulatory compliance, particularly when sensitive or personal information is involved.

To address these challenges, several collaborative ML paradigms have been proposed, among which Federated Learning(FL) \cite{FA0} has emerged as one of the most prominent. In a typical FL setup, participating clients are organized in a star shaped topology, where each client communicates with a central server. Instead of transmitting raw data, the server distributes a global model to clients which then perform local training using their on device data and computation resources. The resulting model updates are sent back to the server, where they are aggregated to produce an improved global model. This iterative process continues until convergence. FL has been successfully adopted in real world applications, such as training the default Android keyboard, G Board \cite{gbw1} demonstrating its practical viability.

This privacy focused learning paradigm, enables resource constrained devices to collaboratively train high quality models, supports large scale learning through aggregation, and decouples model training from direct access to raw data. However, despite these benefits existing FL frameworks exhibit notable limitations. Most implementations rely on a centralized server that holds significant control and privileges over participating clients, making them more suitable for enterprise environments with controlled infrastructures. Additionally, this centralized authority often operates in an opaque manner, offering limited visibility raising concerns over how client updates are processed or weighted.

These challenges highlight the need for an alternative FL architecture that is inherently decentralized, transparent and verifiable. In this context, the emerging Web3 paradigm presents a compelling solution \cite{web3cool}. Web3 technologies built upon blockchain networks and decentralized storage systems such as InterPlanetary File System, embed decentralization, immutability and verifiability at their core. Blockchain enables transparent and tamper proof record keeping, fostering trust among participants, while IPFS provides a distributed and highly available content delivery layer that mitigates single point of failure. Together, these technologies offer a robust foundation for building federated learning systems that are transparent, auditable and resilient.

In this work, we present JiRAIYA,(\textbf{Ji}RAIYA \textbf{R}eputation and \textbf{A}ggregation \textbf{I}ntegrated \textbf{Y}et \textbf{A}nother FL Implementation), a decentralized FL framework that leverages the Web3 stack. The primary contributions of this paper are as follows:

\begin{itemize}
    \item  Design and implementation of an efficient FL system that operates without reliance on an enterprise grade centralized server.
    \item  Introduction of a transparent and decentralized mechanism for detecting and mitigating malicious model updates.
    \item  Implementation of a fallback mechanism to ensure reliable model generation in adversarial or failure scenarios.
\end{itemize}

This paper is structured into six sections. Section \ref{sec2} provides extensive review of related work in the literature. Section \ref{sec3} introduces the prerequisite concepts, techniques and technologies underlying the proposed framework. Section \ref{sec4} describes the system architecture in detail, including participant roles, workflows and key distinguishing features. Section \ref{sec5} presents and analyzes the experimental results. Finally section \ref{sec6} concludes the paper and outlines potential directions for future research. Table 1 summarizes the abbreviations used throughout the paper.

\begin{table}
    \centering
    \begin{tabular}{|l|c|}\hline
 \textbf{Abbreviation}&\textbf{Full Form}\\\hline
        
         CID& Content Identifier\\\hline
         FL& Federated Learning\\\hline
         IPFS&Inter Planetary File System\\\hline
         IC& Influence Calculator\\\hline
         ND& Novelty Detection\\\hline
         OCSVM& One Class Support Vector Machine\\\hline
         SC& Smart Contract\\ \hline
    \end{tabular}
    \caption{Commonly Used Abbreviations}
    \label{tab1}
\end{table}

\section{Literature Review}\label{sec2}
From its inception, numerous optimization techniques have been proposed to make FL more efficient and robust, primarily through the design of improved aggregation algorithms. In order to reduce the communication costs, Konecny et al \cite{FA0} introduced two communication efficient strategies namely, Structure Updates and Sketched Updates. Additionally, building upon the original FedAvg algorithm, several variants have been proposed to address convergence, robustness and heterogeneity, including FedAvgM \cite{FA1}, FedTrim \cite{FA2}, FedProx \cite{FA3}, SCAFFOLD \cite{FA4} and Krum \cite{FA5}.
FL has been widely adopted across diverse application domains, raining in from Internet of Things environments to healthcare and supply chain management. Wazzeh et al proposed a hybrid FL and Split learning framework, termed Dynamic Split Federated Learning to mitigate privacy concerns in IoT settings. Their approach employs genetic algorithms for client selection, ensuring optimal resource utilization during training across multiple heterogeneous services. Badaoui et al \cite{FL4}, presented a federated extension on the existing E-CATBraTS framework for multimodal brain MRO segmentation and proposed an aggregation method called DaQAvg. DaQAvg considers both data size and data quality as key parameters while aggregating and was evaluated against FedAvg,FedProx and FedNova. Zheng et al \cite{FL5} implemented FL for supply risk prediction by clustering participants into collaborative teams. This work enables the participants to join a collective information sharing network, thus enhancing the model performance and convergence rate.

Despite its privacy centric design, FL is not immune to adversarial attacks, particularly due to its collaborative and distributed training paradigm. Malicious actors can disrupt the training process at multiple stages. During local training, attackers may manipulate labels or poison data as discussed by Nowroozi et al \cite {AOF1}. Alternatively, attacks can be launched during the aggregation phase by introducing a large number of fake or compromised clients to poison the global model. Cao et al \cite{AOF2} described one such attack, termed MPAF, where the attackers dragged the global model towards the chosen base model thus affecting the performance of the global model. Membership Inference attacks are another significant threat vector that involves participating clients attempting to infer whether a specific data record was part of an individual training set. Bai et al \cite{AOF3} systematically categorizes these attacks based on their characteristics and provides a comprehensive summary of existing defense strategies. Xia et al \cite{AOF4} classifies the attacker's intent into three categories: Target Semi Targeted, and Untargeted Poisoning attacks. In targeted poisoning attacks, the adversary aims to degrade the performance of the global model on a specific task rather than across all tasks, with backdoor attacks serving as a prominent example. In semi-targeted attacks, the objective is to mis-classify a particular source class while maximizing attacks effectiveness. In contrast, untargeted poisoning attacks aim to broadly degrade the performance of the global model or prevent it from converging altogether.

To mitigate the attacks on FL, numerous defense strategies have been proposed.  Yazdinejad et al \cite{DOF1} introduced a defense mechanism specifically designed to counter model poisoning attacks. Their approach incorporates an internal auditor that evaluates encrypted gradient similarity and its statistical distribution to differentiate between honest and malicious client updates. This distinction is achieved using a Gaussian Mixture model and Mahalanobis Distance, enabling Byzantine tolerant aggregation. Furthermore, the proposed framework leverages additive homomorphic encryption to preserve gradient confidentiality while maintaining low computational and communication overhead. Zang et al \cite{DOF2} proposed a defense strategy aimed at mitigating label flipping attacks in FL systems. This method employs dimensionality reduction techniques, specifically a combination of KPCA and K-means clustering, to detect and filter malicious updates. Along similar lines, Liu et al \cite{DOF3} presented a related approach that relies on PCA for dimensionality reduction couples with data mining techniques to estimate similarity and diversity among client updates. Pati et al  \cite{FL3} reviewed the security threats and corresponding mitigation strategies in FL, with a specific focus on healthcare applications. The authors categorized existing defense mechanism in to Controlled Use Protection and Free Use Protection. Controlled use protection enables model training without direct access to raw data whereas Free Use Protection focuses on limiting the influence of participating entities during inference.
\begin{table}
    \centering
    \begin{tabular}{|p{0.15\linewidth}|p{0.25\linewidth}|p{0.25\linewidth}|p{0.25\linewidth}|} \hline 
         \textbf{Ref}& \textbf{ Approach}&  \textbf{Similarities with our work}& \textbf{Differences with our work}\\\hline
         \cite{DOF1}&  Introduces an internal auditor that evaluates encrypted gradient similarity using Mahalanobis Distance and Gaussian Mixture Models.&  Mitigates model poisoning attacks in FL.& Relies on novelty detection and a consensus mechanism to distinguish between honest and malicious updates.\\\hline
         \cite{DOF2}&  Employs KPCA and K Means Clustering to detect label flipping attacks&  Uses ML and dimensionality reduction techniques for attack mitigation.& Leverages a Hierarchical structure to reduce the impact of attacks rather than fully eliminating malicious updates.\\\hline
         \cite{BCDOF1}&  Utilizes smart contract based aggregation and voting mechanisms to remove the single point of failure&  Enhances transparency through blockchain integration.& Relies on a gas efficient design and limits blockchain to a supporting roles rather than a core component.\\\hline
         \cite{BCDOF2}&  Proposes an incentive driven FL framework combining reputation mechanisms with reverse auction theory&  Evaluates participants based on the quality of their model updates to contributions.& Uses active consensus based mechanisms instead of passive reputation evaluation.\\\hline
         \cite{BCDOF4}&  Presents a blockchain based FL framework to discourage backdoor attacks by detecting malicious behaviour at the end of training.&  Aims to deter malicious participants in FL& Relies on non monetary penalties instead of financial punishments enforced by smart contracts.\\\hline
     
    \end{tabular}
    \caption{Summary of selected related literature}
    \label{tab:lrsummary}
\end{table}
In addition to conventional defense mechanisms, several blockchain based solutions have been proposed to enhance the robustness and trustworthiness of FL. Dong et al \cite{BCDOF1} replaced the centralized FL server with a blockchain infrastructure to mitigate the single point of failure inherent in traditional FL architecture. Their framework employs smart contract based voting along with reward and slashing mechanisms, to detect and deter malicious behavior among participants. Zhang et al \cite{BCDOF2} proposed an incentive driven FL framework that integrates reputation mechanisms with reverse auction theory. In this approach, participant reputation and incentives reflect the quality and reliability of their data. Desai \cite{BCDOF4} et al introduces BLockFLA, a blockchain based FL framework designed to discourage backdoor attacks. Their approach focuses on detecting malicious behavior at the end of the training phase and penalizing attackers through monetary punishments enforced via smart contracts. By integrating detection and punishment mechanisms into the blockchain infrastructure, BlockFLA  aims to incentivize honest participation and enhance the security of collaborative training.

Few selected works are summarized in the table \ref{tab:lrsummary}, highlighting their core approaches, similarities and differences with the proposed work. The literature review reveals a gap in existing approaches where defense mechanisms in FL are often external, or dependent on third party entities. An ideal solution would integrate defense mechanisms directly into the FL training process, eliminating reliance on additional participants. Furthermore, while blockchain based solutions enhance security and trust, existing solutions position it as a central component. Instead, blockchain should serve only as a supporting layer, offloading verification and logging tasks without interfering with the core learning. 
\section{Prerequisites}\label{sec3}

This section outlines the key technologies and methods that form the foundation of the proposed framework.
\subsection{Aggregation Functions in FL:}
Aggregation functions play a crucial role in the FL process. They are responsible not only for aggregating local model updates to enable better global learning but also to mitigate the impact of adversarial or faulty client updates. Let $i$ denote the $i^{\text{th}}$ communication round in Federated Learning.
The global model at round $i$ is represented by $\mathbf{W}_i$.
Let ${\mathbf{U}_i^1, \mathbf{U}_i^2, \ldots, \mathbf{U}_i^K}$ denote the updates received from $K$ participating clients. The global model for the next round, $\mathbf{W}_i+1$ is computed using an aggregation function as described below.
\begin{itemize}
    \item \textbf{\textit{Federated Averaging(Fed Avg):}} Fed Avg, aggregates client updates by computing their average and applying it to the current global model. In many cases, a weighted average is used, where the weights correspond to the number of local training samples used by each client. The update rule is given by:
        \begin{equation}
\mathbf{W}_{i+1} = \mathbf{W}_{i} + \sum_{k=1}^{K} \frac{n_k}{\sum_{j=1}^{K} n_j} \mathbf{U}_{i}^{k}
\end{equation}
where  $n_k$ represents the number of local training samples held by client $k$.

    \item \textbf{\textit{FedAvg with Momentum(FedAvgM): }}FedAvgM extends standard FedAvg by incorporating a server side momentum term to stabilize updates and accelerate convergence. This momentum term is denoted by BETA and updation rule is given by:
    Let $\mathbf{V}_i$ denote the server momentum at round $i$ and $\beta \in [0,1)$ be the momentum coefficient. The aggregated update and global model update are given by:

    \begin{equation}
\mathbf{V}_{i+1}
=
\beta\, \mathbf{V}_{i}
+
\sum\limits_{k=1}^{K}
\frac{n_k}{\sum\limits_{j=1}^{K} n_j}
\mathbf{U}_{i}^{k}
\end{equation}
where  $n_k$ represents the number of local training samples held by client $k$.
    \begin{equation}
    \mathbf{W}_{i+1} = \mathbf{W}i + \mathbf{V}{i+1}
    \end{equation}
    \item \textbf{\textit{FedTrim:}} FedTrim is a robust aggregation approach that discards a fraction of client updates deemed to be extreme or outliers before aggregation. Let $\mathcal{S}_i \subseteq {1, \ldots, K}$ denote the set of non-trimmed client indices after discarding extreme updates.. The aggregation rule is defined as:

    \begin{equation}
\mathbf{W}_{i+1}
=
\mathbf{W}_{i}
+
\frac{1}{|\mathcal{S}_i|}
\sum\limits_{k \in \mathcal{S}_i}
\mathbf{U}_{i}^{k}
\end{equation}
    \item \textbf{\textit{Krum:}} Krum is a Byzantine robust aggregation algorithm that selects a single representative update instead of averaging all updates. This approach assumes that honest client updates are relatively close to each other in parameter space, while malicious updates appear as outliers. For each client update $\mathbf{U}_i^k$, define the Krum score as the sum of squared distances to its closest neighbors:
\end{itemize}

\begin{equation}
k^* = \arg\min_k \sum_{j \in \mathcal{N}_k} \left\lVert \mathbf{U}_i^k - \mathbf{U}_i^j \right\rVert_2^2
\end{equation}

The global model is updated using the selected update:

\begin{equation}
\mathbf{W}_{i+1} = \mathbf{W}_i + \mathbf{U}_i^{k^*}
\end{equation}

\subsection{Novelty Deteaction:}
Novelty Detection (ND) refers to the process of identifying abnormal or previously unseen patterns within a dataset. ND is also referred to as semi-supervised anomaly detection, as it assumes that the training data is free of outliers and aims to detect whether new observations deviate significantly from the learned distribution. The general approach involves learning an approximate boundary around the normal data by analyzing the distribution of training samples in embedding space. Observations that lie within this boundary are classified as inliers, whereas those outside are considered outliers. According to Pimentel et al \cite{nd}, novelty detection techniques can be broadly categorized into \textit{probabilistic, distance based} and\textit{ reconstruction based} methods. \textit{Probabilistic ND} estimates the generative probability density function of the data and applies a threshold to determine novelty. As the name suggests, \textit{Distance based ND} identifies outliers based on their  distance from neighboring points using clustering or nearest neighbor approaches; points that are far from their neighbors are classified as outliers.\textit{ Reconstruction based ND} relies on the algorithm's ability to reproduce the input instance; those observations that have high reconstruction error rate are considered as anomalies.

Popular algorithms for ND include \textit{Local Outlier Factor} (LOF) \cite{ND0} and \textit{One Class Support Vector Machine(OCSVM)} \cite{ND1}. In this work, OCSVM is employed due to the sensitivity of LOF to hyperparameter selection. OCSVM is a kernel based method derived from standard SVMs. It constructs a hyperplane that maximizes the distance from the origin while separating inliers from outliers. An alternative formulation by Tax et al \cite{ND2} uses a hypersphere instead of hyperplane to implement ND. In both implementations, the algorithm strives to maximize the margin between inliers and outliers, providing a robust framework for detecting previously unseen anomalies.

\subsection{Blockchain:}
Stauart Haber and W Scott Stornetta \cite{BC1} first proposed a cryptographic mechanism to prevent the manipulation of historical records by future entities, thereby creating tamper proof record keeping. This concept was later extended by Satoshi Nakamoto through the integration of cryptographic primitives, gate theoretic incentives, and a decentralized consensus mechanism in the design of \textit{Bitcoin }\cite{BC2}. Bitcoin is a decentralized cryptocurrency network with Bitcoin as its native token, enabling peer to peer transfer of value without reliance on centralized authorities such as banks or other financial institutions. Transactions occurring on the Bitcoin network are recorded in blocks, and all participating nodes maintain either a partial or full copy of these blocks. Each block contains the cryptographic hash of its predecessor. Any modification to a previous block propagates changes to all subsequent blocks, making tampering easily detectable. This chaining process establishes blockchain both as a tamper detection mechanism and trust establishment platform. To achieve consensus among the distributed nodes in the network, Bitcoin relies on the Proof of Work mechanism.

Building upon these ideas, Vitalik Butering proposed \textit{Ethereum} \cite{BC4} in late 2013. Ethereum, with Ether as its native token has proposed the revolutionary concept of Ethereum Virtual Machine, which enables the execution of code directly on the blockchain, thereby making it Turing complete. Another key innovation of Ethereum is the concept of Smart Contracts, which are autonomous programs deployed on the blockchain that can execute transactions and enforce predefined rules on behalf of users. Gas is the unit that measures the computational work required to execute operations on the Ethereum network. Smart contracts are typically written in languages such as Solidity or Vyper, and applications that leverage smart contracts for backend logic are commonly referred to as Decentralized Application (DApps).

\subsection{Inter Planetary File System:}
InterPlanetary File System(IPFS) , initially proposed by Juan Benet \cite{IPFS0}, enables developers to build applications on top of a decentralized storage layer without having to manage the underlying complexity of decentralization. IPFS aims to interconnect all the computers under an unified files system. Each node participating in the IPFS network is referred to as a peer and every peer possesses a public private key pair. A peer is uniquely identified by a Peer ID, which is derived from the cryptographic hash of its public key. Unlike traditional location based addressing used in HTTPS, IPFS employs content based addressing.  When a file is uploaded to the IPFS network, it is partitioned into fixed size chunks of 256kB, with each chunk being cryptographically hashed. These hashes are organized into a Merkle Directed Acyclic Graph, where the root node aggregates the hashes of all descendant nodes. This root hash serves as the Content Identifier (CID) for the file. Content based addressing enables efficient data de-duplication and provides intrinsic data integrity verification as any modification to the content results in a different CID.

To publish or retrieve content, IPFS maintains mapping between Peer IDs and CIDs using a Distributed Hash Table. IPFS relies on a modified version of the \textit{Kademlia DHT} \cite{IPFS-1}, allowing the peers to efficiently locate and retrieve content by directly connecting to the peer. This decentralized lookup and retrieval mechanism enables IPFS to function as an efficient publicly accessible content delivery network. Furthermore, considering the public nature of IPFS, where anyone can access the content associated with a given CID, it is a common practice to encrypt data before uploading it to the network to ensure confidentiality.

\subsection{Snowball Consensus Protocol:}
The Snowball protocol is part of the \textit{Snow} family of protocols, which are leaderless and Byzantine Fault Tolerant consensus protocols \cite{SF1}. The snow family consists of Slush, Snowflake and Snowball, with each protocol building upon the previous and Slush serving as the foundational layer. These protocols rely on network subsampling and the principle of metastability to achieve consensus. Unlike deterministic consensus protocols, probabilistic protocols offer several advantages, including the ability to operate without prior knowledge of all participants, efficient communication and higher scalability.
In Snowball protocol, an undecided node begins by randomly selecting a subset of peers and querying them and subsequently updating its own state based on the received responses. The Snowball protocol has the following hyper parameters:
\begin{itemize}
    \item \textbf{\textit{Number of Nodes(n):}} Total number of nodes in the network.
    \item \textbf{\textit{Quorum Size(k):}} Number of nodes queried in each round.
    \item \textbf{\textit{Alpha($\alpha$):}}Minimum number of identical responses from the quorum required to determine the majority opinion.
    \item \textbf{\textit{Beta($\beta$):}} minimum number of consecutive rounds supporting a specific decision required to finalize consensus.
\end{itemize}
During the execution, an undecided node queries its selected quorum of size $k$. Upon receiving response, it performs a majority check against $\alpha$. If a majority is found, the counter associated with the corresponding opinion is incremented. The node updates its decision to the opinion associated with the highest counter values. If the current opinion differs from the previous one, the consecutive counter is rest; otherwise it is incremented. This process repeats until the consecutive counter reaches $\beta$, at which point the node finalizes its decision.

\section{Roles and Framework Workflow}\label{sec4}
\begin{figure}
    \centering
    \includegraphics[width=0.75\linewidth]{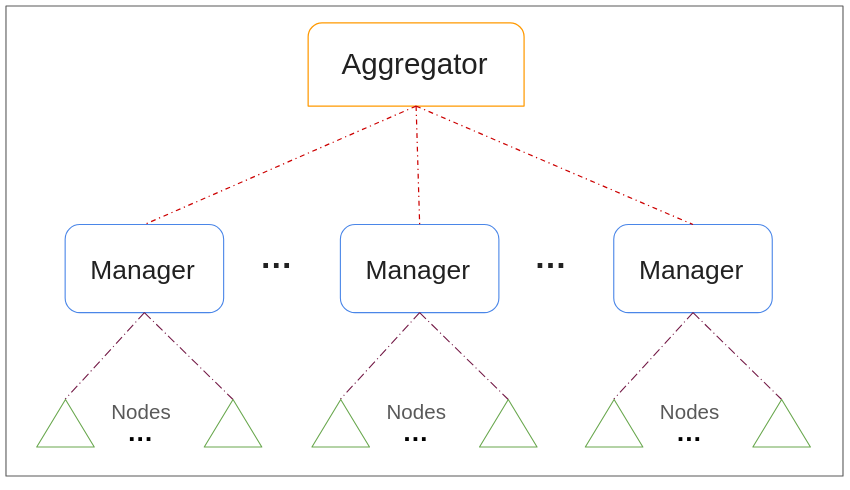}
    \caption{Hierarchical Arrangement of the Roles}
    \label{fig:ar}
\end{figure}

The system architecture is illustrated in Figure \ref{fig:ar}. The proposed framework leverages the IPFS for persistent storage and communication purposes, and employs a smart contract based mechanism for reputation computation, termed \textit{Influence Calculator}(IC).
JiRAIYA is composed of three distinct roles: \textit{Node, Manager and Aggregator}. This section is structured as follows. First, the roles within the proposed platform are described. This is followed by an explanation of the system workflow and the reputation calculation process. Subsequently, the IC is discussed in detail. Finally, the unique features of the proposed framework are presented.

\subsection{Roles:}
JiRAIYA defines three distinct roles organized in a hierarchical structure, with the Aggregator positioned at the top level, Managers at the intermediate level, and Nodes at the bottom. Both the Aggregator and Managers can read from and write to IC.
\begin{itemize}
    \item \textbf{\textit{Nodes:}} Nodes serve as the primary computational units of the proposed framework. They are responsible for training the model with locally available data and are equipped with the required computational resources to support the training process.
    \item \textbf{\textit{Manager:}} Managers are specialized nodes that orchestrate FL within their respective federation. Each node is associated with a single Manager, and each Manager supervises the training process of its assigned nodes. Managers are uniquely identified by a distinct identifier and are responsible for coordinating model aggregation at the federation level and interfacing with the IC.
    \item \textbf{\textit{Aggregator:}} The aggregator is a specialized Manager responsible for coordinating training across all Managers in the proposed framework. Unlike Managers, the Aggregator does not maintain a direct federation of Nodes. Instead, it aggregates the models received from Managers and is responsible for generating the global model. Aggregator is akin to centralized server in conventional to FL settings.
\end{itemize}

\subsection{Workflow}
\begin{figure}
    \centering
    \includegraphics[width=0.99\linewidth]{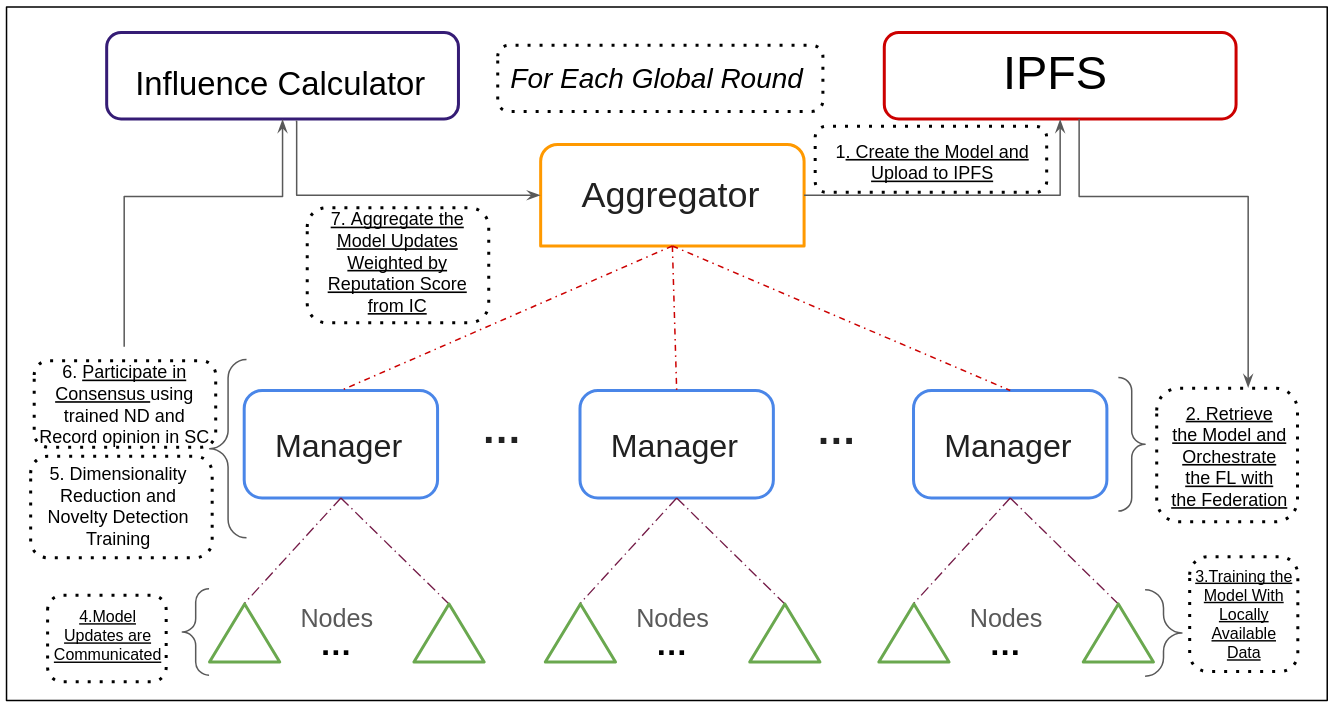}
    \caption{Workflow of the JiRAIYA FL Framework}
    \label{fig:wf}
\end{figure}

The entire training process is divided into three phases: \textit{Training, Consensus and Aggregation}. For each global round, all three phases are executed sequentially. To provide better control over the training process, JiRAIYA introduces three configurable parameters, as illustrated in Figure 1. Local Epochs define the number of epochs each Node performs on its locally available training data. Communication Rounds specify the number of training interactions between Managers and their associated Nodes. Global Rounds determine the number of communication rounds between the Aggregator and the Managers. With each increment in the number of Global Rounds, the number of Communication Rounds increases, while the number of Local Epochs scales proportionally with the Communication  Rounds. The workflow of the proposed framework was illustrated in Figure \ref{fig:wf}.

The \textit{Training phase} constitutes the core of the learning process, with the majority of responsibilities handled by the Managers and Nodes. The process is initiated by the Aggregator, which creates the initial model. In the genesis round, the Aggregator is additionally responsible for training the model before distributing it. Once the model is initialized, the Aggregator communicates the model weights to all Managers. Each Manager retrieves the model weights and orchestrates the FL within its respective federation. Upon receiving the model weights from the Manager, each Node begins training the model using its locally available data for a predefined number of local Epochs. This procedure is repeated for a specified number of Communication Rounds. After the completion of the FL within each federation, the Manager aggregates the model updates received from its associated Nodes. Subsequently, the system transitions to the consensus phase.

The primary objective of \textit{Consensus Phase} is to identify and eliminate malicious or anomalous model updates from the set of proposed updates. During this phase, Managers are responsible for training novelty detection models and evaluating whether the received updates are honest or adversarial in nature.
During the aggregation of model updates, each Manager trains a OCSVM based novelty detection model on the received updates from the nodes. To mitigate potential privacy concerns associated with training directly on raw model updates, the updates are first transformed into lower dimensional representations using Principal Component Analysis(PCA). These PCA based representations are subsequently used to train the novelty detection model. Following this, the model updates proposed by each Manager are collectively assessed by all other Managers. A snowball based consensus protocol is employed to determine whether a proposed update should be accepted or rejected. Upon completion of the Snowball consensus process, each Manager's individual opinion, the corresponding PCA coverage, and the final consensus outcome are recorded in IC. This marks the conclusion of the consensus phase.

In the \textit{Aggregation Phase}, the Aggregator is responsible for generating the aggregated model to be used in the subsequent training round. The aggregated model is produced by combining the accepted model updates with equal weighting. In scenarios where no proposed model updates are accepted or where all proposed updates are accepted, aggregation is performed based on the reputation scores of the contributing entities.

For each proposed model update, each manager's individual opinion is compared against the consensus opinion. If the two opinions match, the corresponding manager's reputation score is incremented by a predefined value; otherwise, the score is detected by the same value. This procedure is applied iteratively to all the managers. To prevent negative values, any negative scores are clipped to minimum values. For transparency, the reputation computation is performed by the IC and was invoked by the Aggregator. The resulting reputation score is ten multiplied by the PCA coverage of each manager. The PCA coverage incorporates a temperature based transformation that emphasizes managers with higher PCA coverage. This design choice ensures that high quality updates receive greater weights as reward for honest behavior. Finally, the product of the reputation score and the PCA coverage is normalized. The resulting product is used as the weightage in the aggregation function.

This sequence of Training, Consensus and Aggregation Phase are implemented in a single round. The entire process is repeated iteratively until the predefined number of Global Rounds are reached

\subsubsection{Reputation score Example:}

\begin{table}
    \centering
    \begin{tabular}{|l|c|c|c|c|c|}\hline
         &  \(U_1\)&  \(U_2\)&  \(U_3\)&  \(U_4\)&  \(U_5\)\\\hline
         \(M_1\)&  1&  0&  1&  1&  0\\\hline
         \(M_2\)&  0&  1&  0&  0&  1\\\hline
         \(M_3\)&  0&  0&  1&  0&  0\\\hline
         \(M_4\)&  0&  0&  0&  1&  1\\\hline
         \(M_5\)&  1&  1&  0&  0&  0\\\hline 
 \(C\)& 0& 0& 0& 0& 0\\ \hline
    \end{tabular}
    \caption{Manager's Individual Opinion }
    \label{tab:op}
\end{table}
For a given global round \( x \), let \( M_1, M_2, M_3, M_4, M_5 \) denote the set of managers, and let \( U_1, U_2, U_3, U_4, U_5 \) represents the mode updates proposed by managers \( M_1, M_2, M_3, M_4, \) and \( M_5 \), respectively. Let C denote the final consensus opinion for the round. 
Each entry in Table \ref{tab:op} represents the opinion expressed by the corresponding manager on the proposed model update for global round \( x \). Since none of the proposed updates are accepted, the IC based mechanism is invoked. Let the reputation score updates with fixed increment and decrement values of 5, determined by matching operation between each manager's individual opinion and the consensus opinion. Specifically, if a manager's opinion is matched with  \(C \), its reputation score is increased by 5; else it is penalized by 5.

\begin{table}
    \centering
    \begin{tabular}{|l|c|c|c|c|c|}\hline
         &  \(M_1\)&  \(M_1\)&  \(M_1\)&  \(M_1\)& \(M_1\)\\\hline
         \textbf{Reputation Score from IC}&  1&  5&  15&  5& 5\\\hline
         \textbf{PCA Coverage}&  35&  50&  75&  55& 60\\\hline
         \textbf{Transformed PCA Coverage}&  0.087&  0.149&  0.370&  0.179& 0.215\\\hline
         \textbf{Final Weightage}&  0.018&  0.126&  0.567&  0.139& 0.151\\ \hline
    \end{tabular}
    \caption{Final Weightage of Managers}
    \label{tab:we}
\end{table}
The IC derived reputation scores and the PCA coverage values for each manager are summarized in table \ref{tab:we}. Using a temperature parameter of 0.1, the PCA coverage values were transformed accordingly. These transformed PCA values were then multiplied by the corresponding reputation scores, and the resulting products were normalized to obtain the final aggregation weights reported in the table. As mentioned in the previous section, the aggregation was weighted by the Final Weights.

\subsection{Influence Calculator}
The Influence Calculator (IC) constitutes the core component of the proposed platform and is implemented in Solidity. It is designed with a string emphasis on gas efficiency and minimal attack surface. To achieve this, the IC incorporates roles based access control at the contract level, ensuring that each function can only be invoked by authorized roles. Write access to the IC is permissioned, requiring Managers to register by making deposits. Each Manager is uniquely identified by an individual identifier, as mentioned in the previous section. The contract relies heavily on Solidity mappings, which provide an efficient key value storage mechanism and contribute significantly to the contract's performance.

The IC maintains several mappings to manage participation, record opinions, and to calculate reputation score, throughout the training lifecycle. The \textit{managerOpinion} mapping is a nested mapping that stores the opinions submitted by Managers across all training rounds. Specifically, the global round number serves as the primary key, mapping to another nested mapping that associates each Manager's unique identifier with their submitted opinion. This structure enables efficient retrieval and aggregation of opinions on a per round basis.  To enforce access control, the\textit{ validManager }mapping is used to verify the legitimacy of a Manager. It maps a Manager's public address to a boolean value, where a value of true indicates a registered and authorized participant. This mapping is directly leveraged by access control modifiers to restrict function execution.
\begin{figure}

\label{e1}
\begin{subfigure}{0.45\textwidth}
    
    \includegraphics[width=\textwidth]{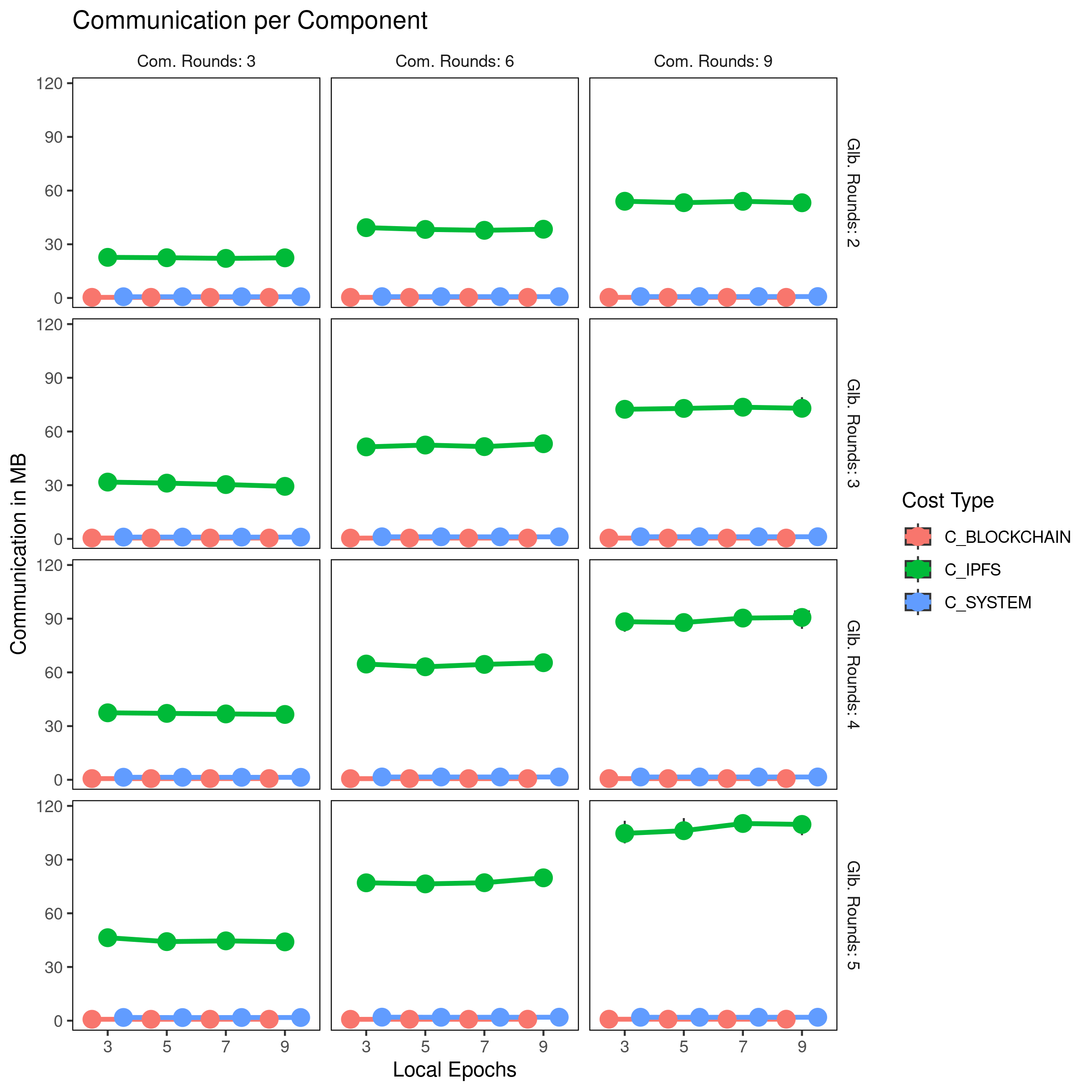}
    \caption{Communication Overhead of JiRAIYA}

    \label{comm}
\end{subfigure}
\hfill
\centering
\begin{subfigure}{0.45\textwidth}
    
    \includegraphics[width=\textwidth]{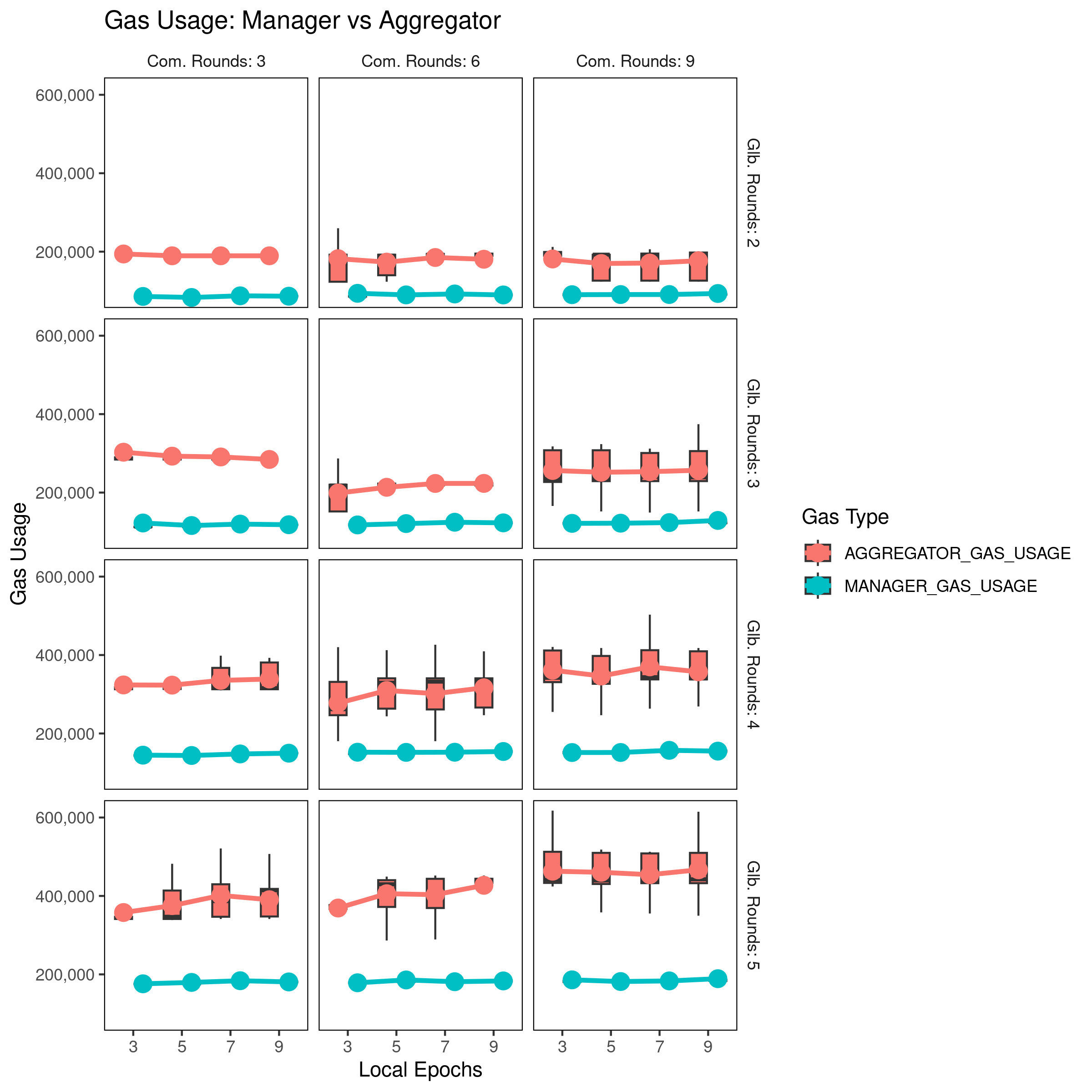}
    \caption{Gas Usage of Aggregator and Manager}
    \label{gu}
\end{subfigure}
\hfill

\label{acc}
\begin{subfigure}{0.65\textwidth}
    \includegraphics[width=\textwidth]{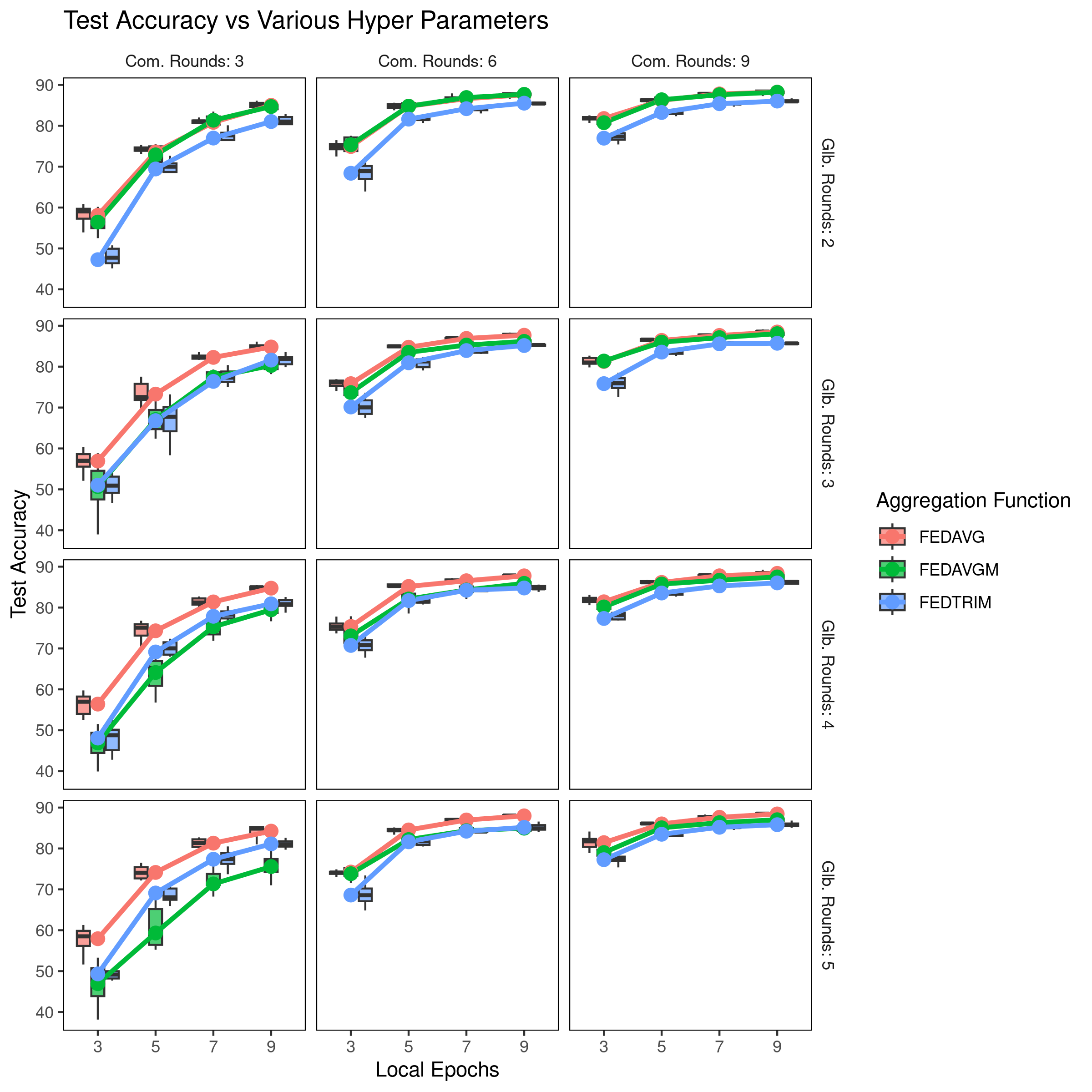}
    \caption{Communication Overhead of JiRAIYA}

    \label{exp1}
\end{subfigure}

\caption{Gas, Communication Usage and Test Accuracy of JiRAIYA.}
\end{figure}

The \textit{pcaCoverage} mapping associates each Manager's unique identifier with their corresponding PCA coverage value, which is provided during opinion submission and later used in reputation evaluation. The \textit{consensusOpinion} mapping the round number to a string representation of the final consensus opinion. Finally, the \textit{reputationScore} mapping stores the computed reputation score for each Manager, mapping the unique identifier.

Furthermore, the IC defines two primary modifiers to regulate function execution: \textit{onlyValidManager }and \textit{onlyAggregator}. The \textit{onlyValidManager} modifier restrict access to register Managers by consulting the \textit{validManager} mapping, effectively distinguishing between authorized participants and external entities. The \textit{onlyAggregator} modifier ensures that only the designated Aggregator, identified through the \textit{AGGREGATOR} state variable, can invoke specific privileged functions. These modifiers significantly reduce the contract's attack surface by enforcing strict role separation.
\begin{figure}[!ht]

\label{e1}
\begin{subfigure}{0.45\textwidth}
    
    \includegraphics[width=\textwidth]{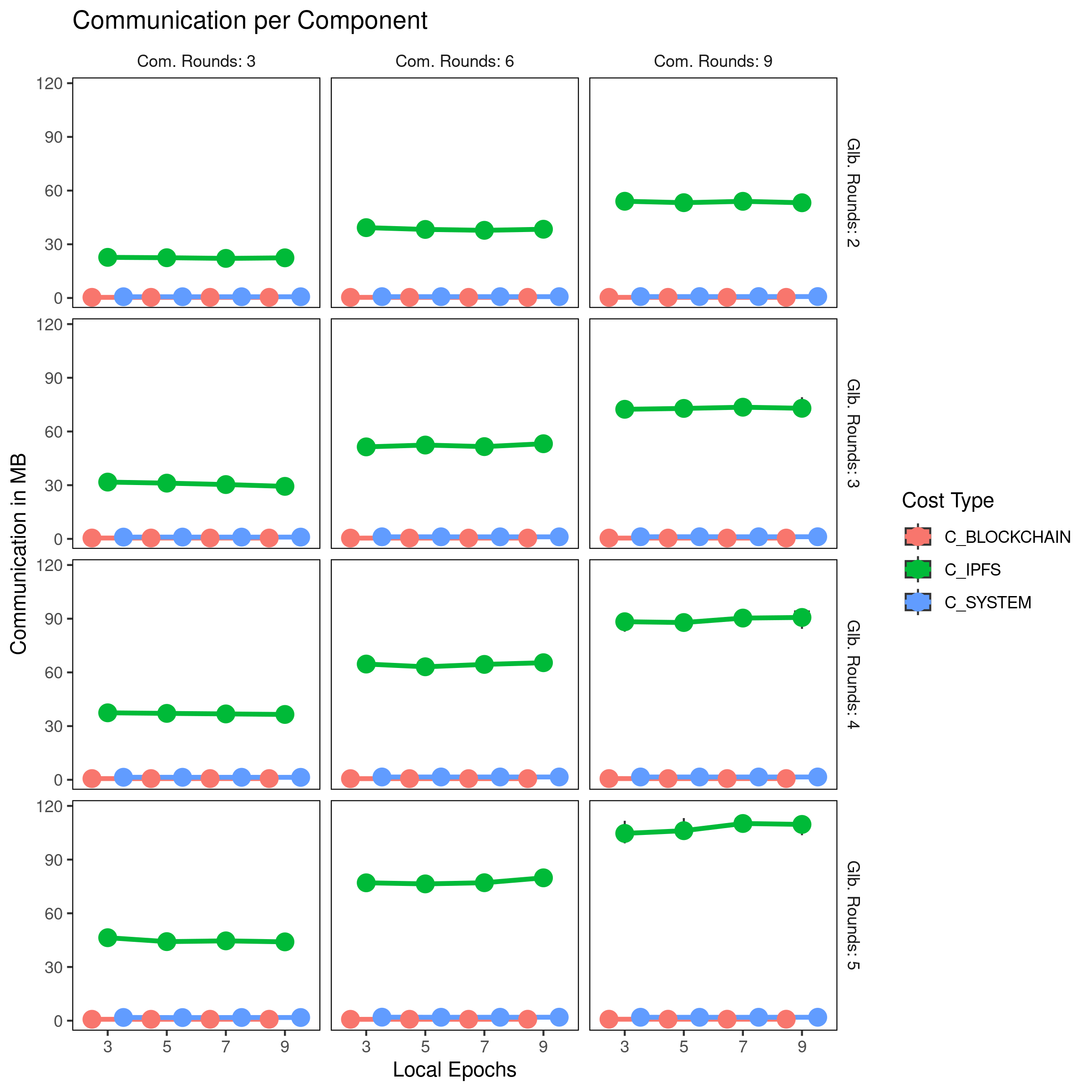}
    \caption{Communication Overhead of JiRAIYA}

    \label{comm}
\end{subfigure}
\hfill
\centering
\begin{subfigure}{0.45\textwidth}
    
    \includegraphics[width=\textwidth]{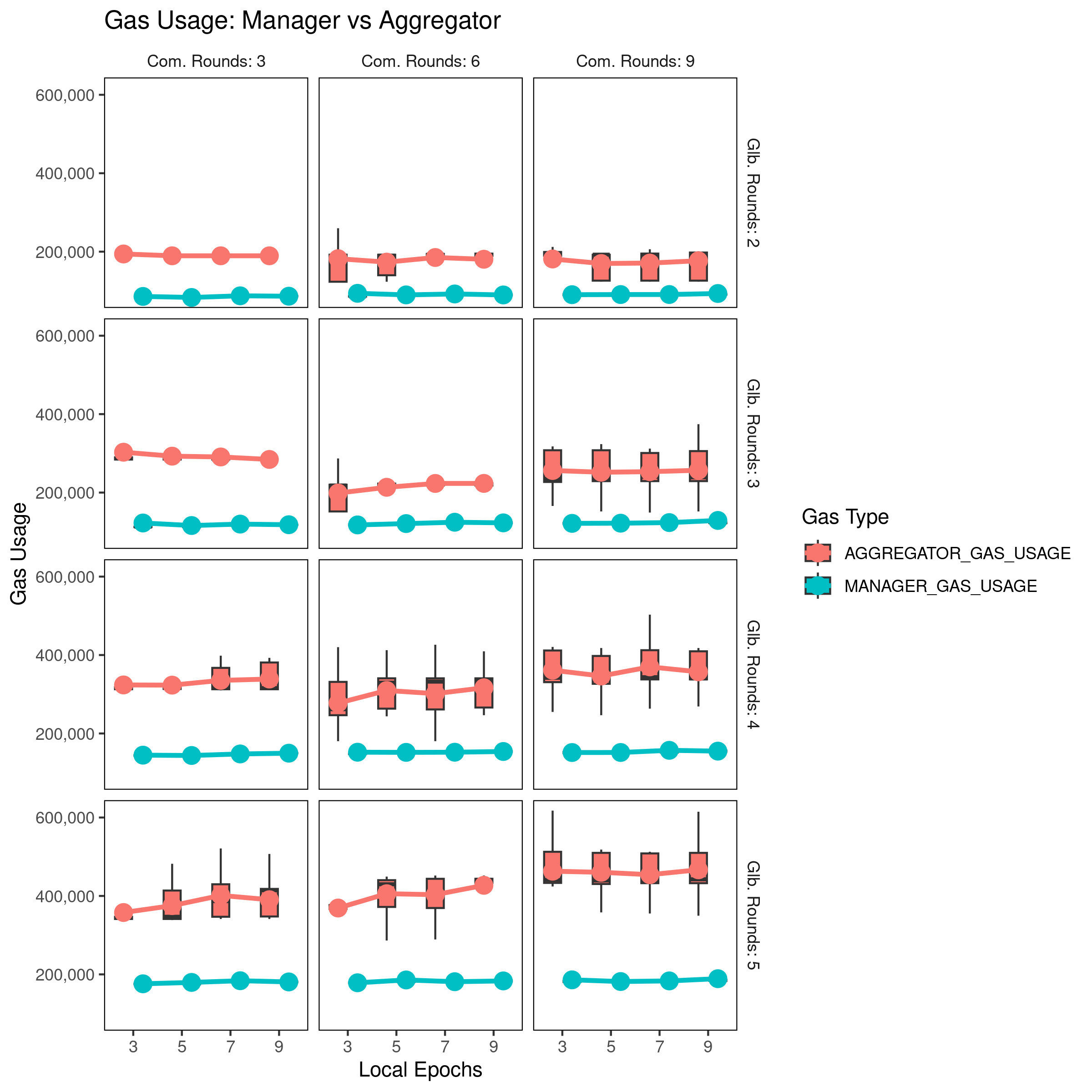}
    \caption{Gas Usage of Aggregator and Manager}
    \label{gu}
\end{subfigure}
\hfill

\label{acc}
\begin{subfigure}{0.65\textwidth}
    \includegraphics[width=\textwidth]{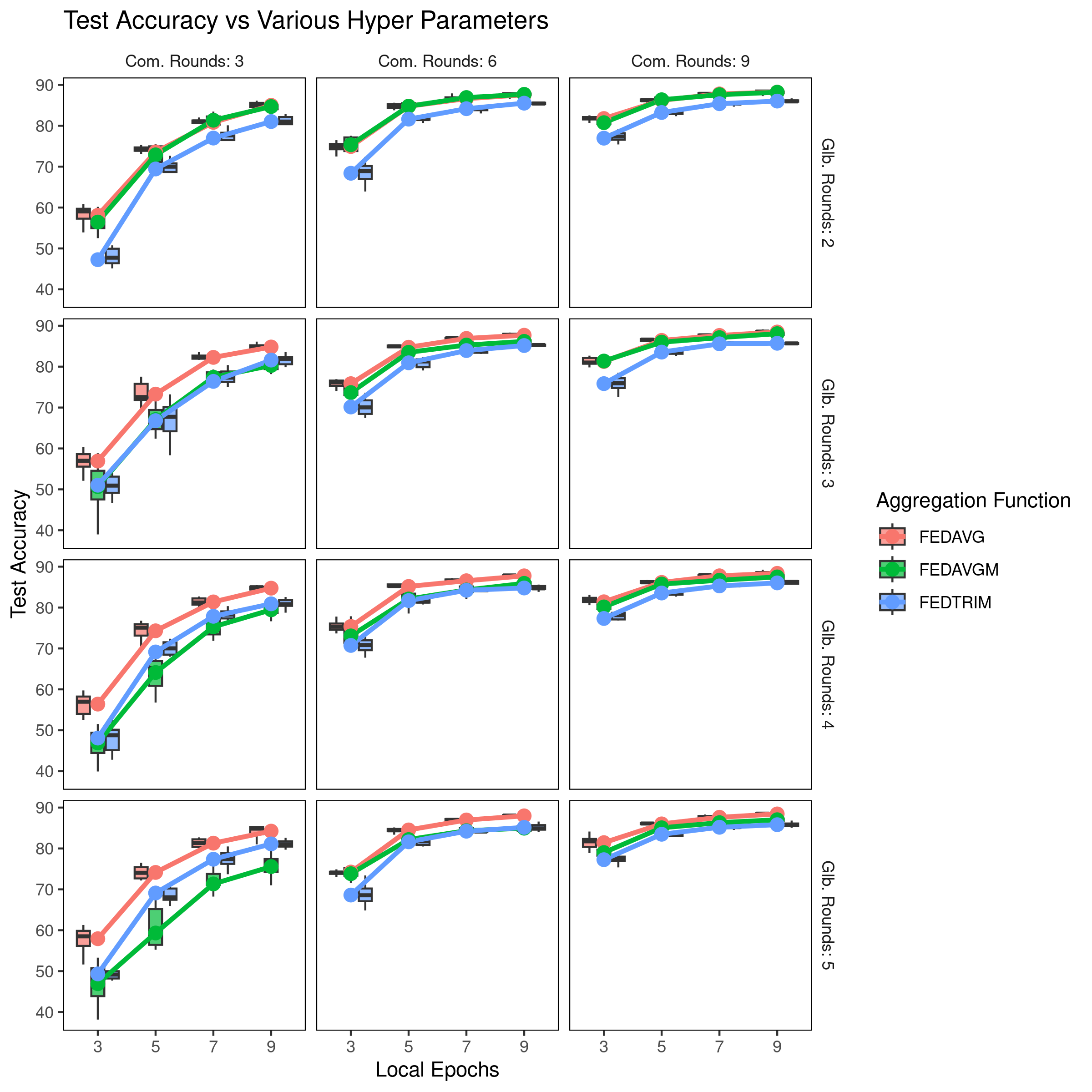}
    \caption{Communication Overhead of JiRAIYA}

    \label{exp1}
\end{subfigure}

\caption{Gas, Communication Usage and Test Accuracy of JiRAIYA.}
\end{figure}
IC relies on 3 core functions, namely \textit{registerManager} function enables any entity intending to participate as a Manager to join the framework. This function is payable in nature and requires the caller to submit a minimum deposit. It utilizes \textit{`msg.sender'} to identify the registering address and \textit{`msg.value'} to verify that the deposited amount meets the required threshold. Upon successful validation, the function updates the \textit{validManager} mapping, granting the caller write access to Manager restricted functions. The \textit{addManagerOpinion} function is restricted to valid Managers through the \textit{onlyValidManager }modifier. This function records a Manager's opinion for a given global training round by updating the \textit{managerOpinion} mapping. It accepts the current global round number, the Manager's unique identifier, the individual opinion on the model updates in string format, and the associated PCA coverage value, which is simultaneously stored in the \textit{pcaCoverage} mapping. 

The\textit{ computeReputationScore} function can only be invoked by the Aggregator. It computes the reputation score for each Manager by comparing the individual opinions stored in the \textit{managerOpinion} mapping against the corresponding consensus values recorded in the \textit{consensusOpinion} mapping. The resulting score is clipped to enforce a minimum value of one and the score's increment and decrement is set to 5. The final computed values are then stored in the \textit{reputationScore} mapping, ensuring persistent tracking of Manager reliability over time.

\subsection{Unique Features}
This section explores the unique features of the proposed framework.
\subsubsection{Hierarchical Role Based FL Architecture:}

JiRAIYA introduces a hierarchical, role based FL architecture that eliminates the dependency on enterprise centric coordination models. By delegating orchestration responsibilities to intermediate Managers and restricting the number of model updates processed by each Aggregator, this framework prevents the aggregator from becoming a performance or security bottleneck. This hierarchical decomposition significantly reduces the influence of malicious or anomalous updates on the trained model. Moreover, the role based design facilitates the fine grained access control and permission management in IC.

\subsubsection{ND and Consensus Driven Mechanism}

JiRAIYA achieves competitive model performance while maintaining low communication overhead, making it well suited for large scale FL deployments. The framework uniquely integrated ND based validation with consensus mechanisms to accurately identify and filter malicious model updates. This combined defense strategy is highly scalable and enables near instantaneous decision making during aggregation. Importantly, the entire pipeline is integrated into the framework, eliminating reliance on third part tools or external verification platforms.

\subsubsection{Privacy Preserving Efficient FL Implementation:}

During the consensus phase, all the model updates are represented using their PCA representations, ensuring that sensitive parameters are not directly exposed during the validation process while still enabling the effective detection of anomalous or adversarial behavior. When no model updates are accepted, a reputation based fallback mechanism is used to generate the model. This backup mechanism is invoked only when required, thereby minimizing unnecessary computational and gas overhead. Additionally, sensitive parameters like model updates are protected through encryption using a combination of RSA and Fernet cryptographic schemes.

\subsubsection{Web3 Enabled FL Implementation:}

JiRAIYA is fully implemented within a Web3 ecosystem, leveraging blockchain technology for immutable record keeping and IPFS for decentralized content delivery and persistent storage. The framework is designed to operate seamlessly across both public and private blockchain infrastructures with minimal or no architectural modifications. Importantly, interactions among the various roles within the proposed framework are conducted through the exchange of IPFS based CIDs. Furthermore, the IC has minimal attack surface, and does not store sensitive information. Reputation scores are computed directly within the smart contract, enabling transparent and verifiable validation while preserving system security.

\section{Evaluation}\label{sec5}
This section outlines, the implementation details and experimental setup of JiRAIYA, followed by the experimental design and concludes with a discussion of the results.
\subsection{Experiment Set Up}
JiRAIYA was implemented entirely in Python, while the IC was developed using the Foundry \cite{IEF1} toolkit for smart contract development and deployment in a local environment. Interaction with the IC and various roles were facilitated using the Web3py library\cite{IEF2}. A Kubo based daemon was employed to run IPFS with all interactions performed via its HTTP API. Communication among the different roles within the proposed framework was realized through a socket based event driven, bidirectional communication later implemented using Python Socket IO\cite{IEF3}. Encryption of the model updates and other sensitive parameters was done using combination of RSA and Fernet. Model training was carried out using the SciKit- Learn libraby\cite{IEF4}. Each experiment was repeated 10 times to mitigate randomness.

All experiments were conducted on a workstation equipped with an Intel(R) Core(TM) i7-6700T CPU clocked at 2.80 GHz, 16 GB of DDR4 memory, running Pop!\_OS 22.04 LTS in local setting. The proposed framework was trained using a hierarchical configuration of one aggregator and five managers, with each manager supervising three nodes, resulting in a total of 15 participating nodes. No client selection strategies were employed, instead all nodes participated in FL training. The snowball consensus protocol was configured with parameters $k$, $\alpha$, $\beta$ set to 4,3 and 4. These values were empirically evaluated and selected as they yielded the highest probability of achieving single decree across managers. To safeguard model parameters and reduce dimensionality, all model updates were transformed using PCA, retaining the top five principal components.

The MNIST dataset was used to  evaluate the proposed framework. Rather than using the raw image representations, all pixel values were converted to grayscale intensities, where 0 denotes black and 255 denotes white. Logistic Regression model was employed for evaluation, with test accuracy used as the primary performance metric. Data sampling across nodes was performed randomly without replacement.

To study the behaviour of the framework under different aggregation stratgies, three aggregation functons were considered: \textit{Federated Averagining(FedAvg)}, \textit{FedAvg with momentum(FedAvgM)} and \textit{FedTrim}. For FedAvgM, the learning rate was set to 1 and the momentum  coefficient was set to 0.9. For FedTrim, the trimming parameter was set to 0.2, meaning that 20\% of the extreme model update values were discarded during the aggregation. 
It is important to note that the aggregation functions considered are evaluated exclusively at the upper level of the hierarchical architecture, namely between the Manager and the Aggregator. These functions are not applied at the lower level between the Nodes and the Managers. At the node manager interface, standard FedAvg is consistently employed for model aggregation.

\subsection{Experimental Design}
To evaluate the proposed framework under realistic adversarial settings, three experiments were conducted to assess system performance and robustness against real world attacks. In the first experiment, the objective was to identify optimal system parameters, namely Global Rounds, Communication Rounds and Local Epochs. The number of Communication Rounds was varied from 3 to 9 with a step size of 3, Local Epochs were incremented from 3 to 9 with a step size of 2 and Global Rounds were varied between 2 and 5. Since JiRAIYA introduces additional communication overhead due to interactions with IPFS, the blockchain later and inter system roles, the corresponding communications costs were measured alongside the gas consumption incurred under each configuration. IPFS communication captures the volume of data written to and read from IPFS, blockchain communication measures the data exchanged with the blockchain and system communication accounts for the overhead incurred during coordination of the training process and the consensus phase. Additionally, the Gas Usage of Manager and Aggregator was also measured. Based on these results, the most optimal parameter settings were selected and used consistently in the subsequent experiments.

The second experiment focused on evaluating resilience against data poisoning attacks. In this setting, the training data of compromised nodes was manipulated using two approaches: \textit{label flipping}, where original class labels were prelaced with random values, and \textit{feature manipulation}, where noise was injected into the original feature values. The third experiment evaluated the impact of model poisoning attack on the final aggregated model, where adversarial behavior was introduced by modifying model parameters through sign flipping and addition of Gaussian noise.

To quantitatively measure the impact of adversarial behavior, two parameters were introduced: \textit{Flip} and \textit{Proportion}. Flip represents the number of compromised Managers participating in the training process and range from 0 to 4 in increments of 1. Proportion (Prop) represents the intensity of the attack and is defined in the multiples of 25\%, with values ranging from 1 to 4. For example, a configuration of Flip and Proportion set to 2 and 3 indicates the following. 2 out of 5 Managers are malicious with an attack intensity of 75\%. In the case of data poisoning attacks, this implies that 75\% of the raw training data held by the compromised Manager is poisoned through both label flipping and feature manipulation, whereas in the case of model poisoning attacks, 75\% of the model parameters are modified.

The impact of these attacks was compared with the Krum aggregation strategy implemented using the Flower framework \cite{beutel2020flower} with 15 participating clients. To ensure fairness in comparison, the number of Local Epochs were matched between the JiRAIYA and Flower implementations, and all clients were required to participate in every training round. Although this comparison is not strictly equivalent in terms of system architecture, it is important to note that Krum receives model updates directly from all participating clients and selects a single update for aggregation, where as the Aggregator in JiRAIYA receives at most five aggregated updates from the Managers, owing to its hierarchical structure.

\subsection{Discussion}

The results of the first experiment are illustrated in Figure 3. Figure \ref{comm} presents the network communication costs associated with different parameters configurations, Figure \ref{gu} depicts the corresponding gas consumption, and Figure \ref{exp1} reports the test accuracy across these configurations. As illustrated in Figure \ref{comm}, IPFS accounts for the highest network usage among the three communication types. This overhead increases with both Communication Rounds and Global Rounds with more pronounced improvement observed across Communication Rounds than Global Rounds. This behaviour can be attributed to the size of the encrypted payloads and the increased number of exchanges between system toles as Communication Rounds increase, leading to higher IPFS traffic. A similar trend is observed from blockchain communication, where the communication cost increases with the number of Global Rounds while remaining relatively stable across different Communication Rounds. This can be explained by the increasing length of opinions written to the blockchain as training progresses. System communication also increases with both Communication Rounds and GLobal Rounds; however, Global Rounds have more significant impact than that of Communication Rounds.

Gas consumption results are illustrated in Figure \ref{gu}, where gas usage is reported for both the Aggregator and the Manager. The Manager gas usage represents the mean gas usage across all Managers. Across all configurations, the Aggregator consistently incurs higher gas costs than the Managers. This is primarily due to the Aggregator executing gas intensive operations such as reputation score computation, whereas Managers are limited to updating their opinions. Furthermore, gas usage for both roles increases with the number of Global Rounds, while Communication Rounds and Local Epochs exhibit minimal impact on gas consumption.

The test accuracy trends are presented in Figure \ref{gu}. An increase in Local Epochs generally leads to improved test accuracy. The performance difference across varying Local Epoch values are more pronounced at lower Communication Rounds and gradually plateaus as Communication Round increases. Across different values of Global Rounds, only marginal improvements in accuracy are observed, accompanied by a noticeable increase in communication overhead. With respect to aggregation strategies, FedAvg and FedAvgM exhibit comparable performance, whereas FedTrim shows slightly degraded accuracy, which can be attributed to trimming of model updates.

Considering the trade-offs between communication overhead, gas consumption and model performance, the number of Communication Rounds was fixed at 6 and the number of Local Epochs was set to 7 for subsequent experiments. Although Local Epochs could be increased up to 9, higher values were avoided due to their potential to amplify the impact of malicious participants. For Global Rounds. Both values of 2 and 3 yielded similar performance and were therefore retained for further evaluation.

\begin{figure}

\label{exp2}
\begin{subfigure}{0.45\textwidth}
    
    \includegraphics[width=\textwidth]{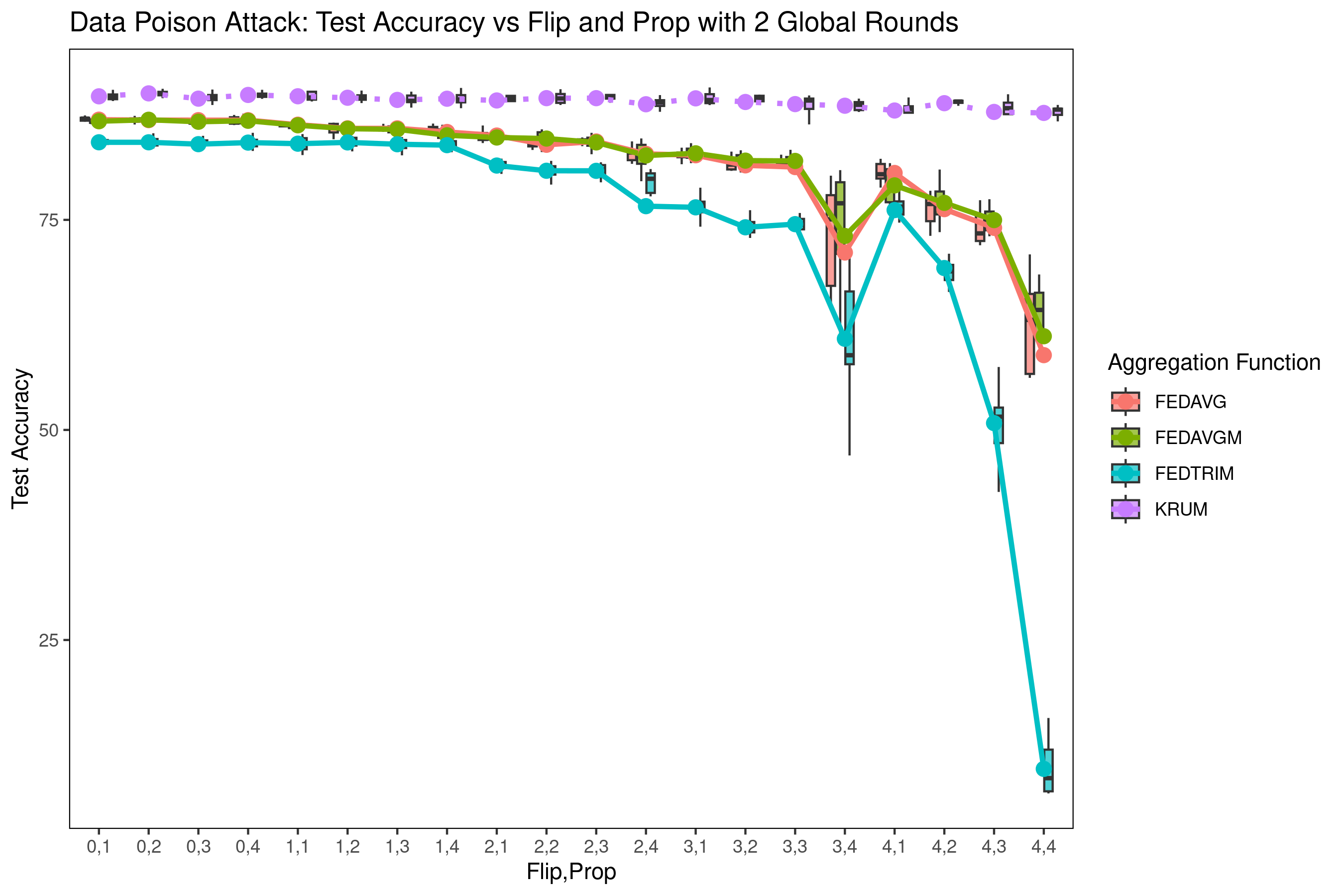}
    \caption{Data Poisoning with 2 Global Rounds}

    \label{e22}
\end{subfigure}
\hfill
\centering
\begin{subfigure}{0.45\textwidth}
    
    \includegraphics[width=\textwidth]{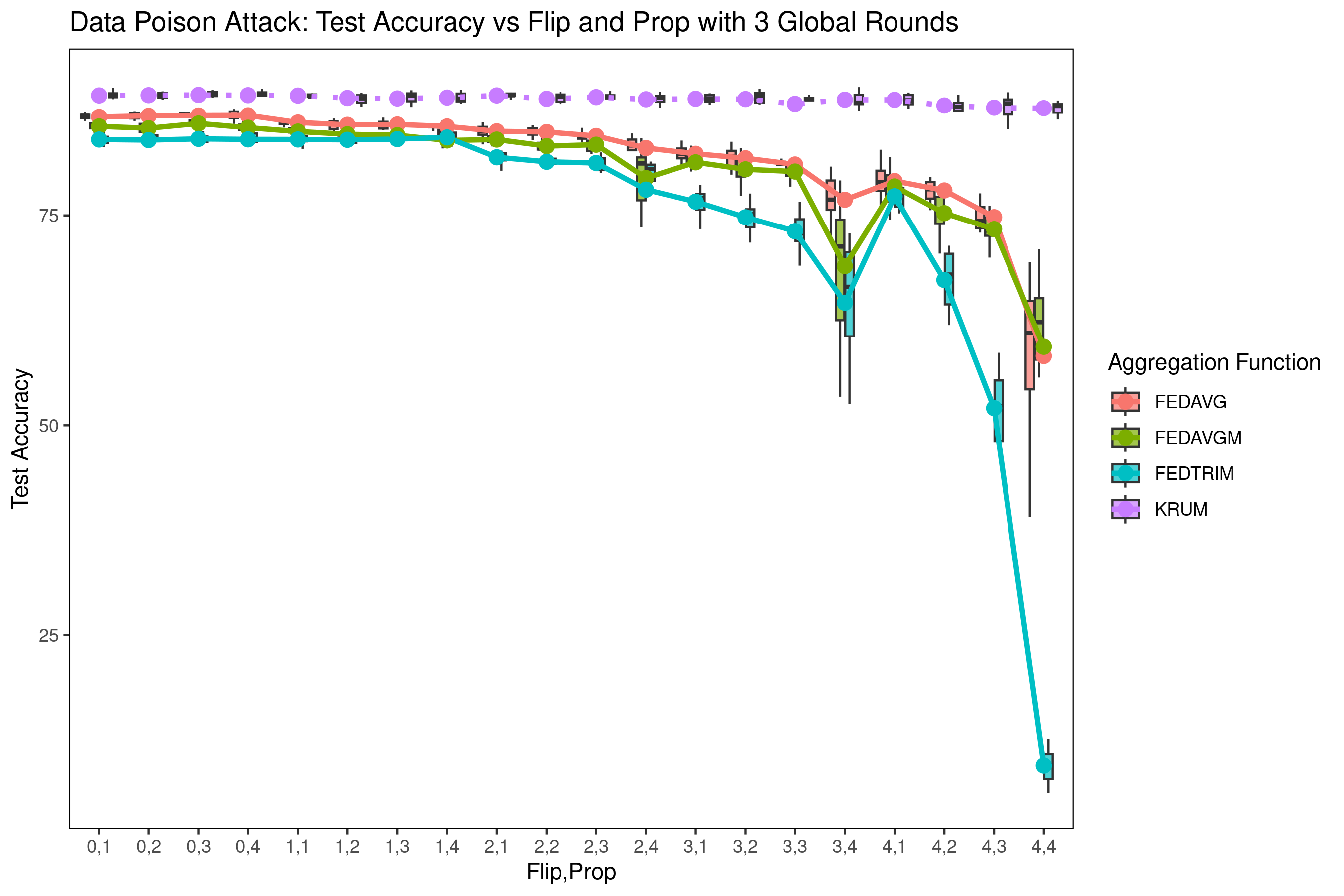}
    \caption{Data Poisoning with 3 Global Rounds}
    \label{e23}
\end{subfigure}
\hfill

\caption{Data Poisoning Attacks with Varying Degree of Flip, Proportion.}
\end{figure}
Figure 4 illustrates the impact of data poisoning attacks on test accuracy and compares the proposed framework with a single level Krum implementation using the Flower framework. Figure \ref{e22} corresponds to Global Rounds set to 2, while Figure \ref{e23} corresponds to the Global rounds set to 3. From Figure 4, increasing the Flip parameter leads to a degradation in performance, with the decline becoming more pronounced at higher Flip values. This degradation in performance, with the decline becoming more pronounced at higher Flip values. In case of FedAvg and FedAvgM, the decline in performance is more pronounced beyond the (Flip, Proportion) configurations of (3,1) whereas beyond (2,1) for Fed Trim. FedAvg and FedAvgM demonstrate similar robustness, with FedAvgM marginally outperforming FedAvg, whereas FedTrim exhibits substantially poorer performance. In contrast, Krum remains largely unaffected by the variations in both Flip and Proportion, maintaining stable performance.
\begin{figure}

\label{e30}
\begin{subfigure}{0.45\textwidth}
    
    \includegraphics[width=\textwidth]{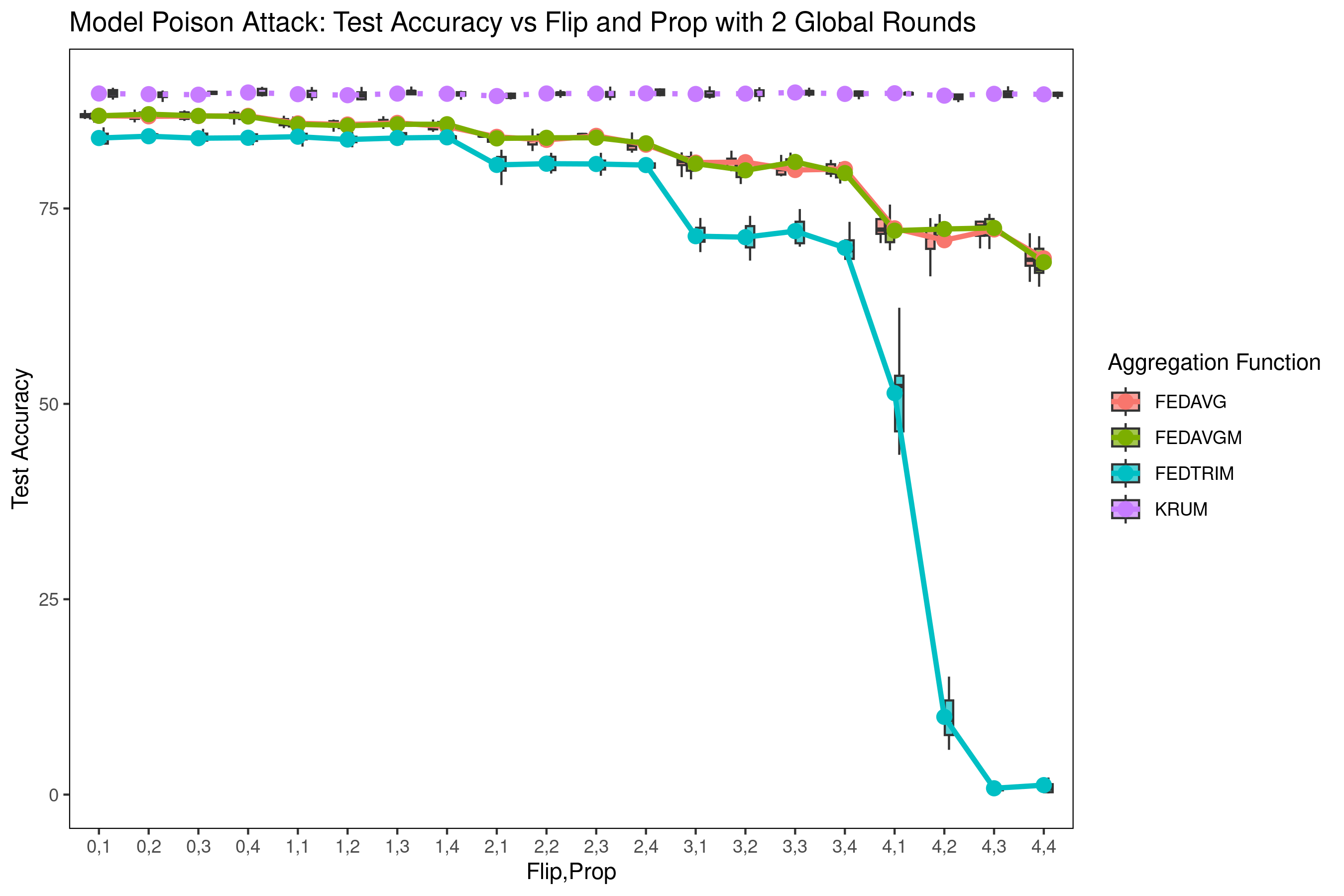}
    \caption{Model Poisoning with 2 Global Rounds}

    \label{e32}
\end{subfigure}
\hfill
\centering
\begin{subfigure}{0.45\textwidth}
    
    \includegraphics[width=\textwidth]{EXP2_GB_2.png}
    \caption{Model Poisoning with 3 Global Rounds}
    \label{e33}
\end{subfigure}
\hfill

\caption{Model Poisoning Attacks with Varying Degree of Flip, Proportion.}
\end{figure}

Figure 5 presents the results for model poisoning attacks, where Figure \ref{e32} corresponds to Global Rounds set to 2 and Figure \ref{e33} corresponds to Global Rounds set to 3. As Flip increases a clear and rapid degradation in test accuracy is observed, indicating the stronger impact of model poisoning compared to data poisoning. The effect of increasing Proportion becomes more pronounced at higher Flip values. For Global Rounds set to 2, FedAvgM slightly outperforms FedAvg, whereas for Global Rounds set to 3, FedAvg demonstrates better performance than FedAvgM. In both cases, FedTrim consistently yields inferior results. Notably the steep performance drop observed when Flip equals 4 corresponds to a scenario where no honest Manager remains in the system.

Overall the experimental results suggest several important insights. Increasing the number of Global Rounds does not necessarily guarantee improvements in test accuracy and instead incurs a substantial increase in communication overhead. In contrast, Local Epochs can be carefully tuned to enhance performance, particularly for more complex datasets. Gas consumption is observed to scale more significantly with Global Rounds that with Communication rounds, further emphasizing the cost associated with the Global Rounds. Notably, configurations with two Global rounds achieve performance comparable to those with three global rounds, indicating diminishing returns beyond certain points. The results also demonstrate that model poisoning attacks are considerably more detrimental than data poisoning attacks. When compared to Krum, the proposed framework achieves comparable performance up to a certain threshold of adversarial presence, beyond which degradation becomes evident. This threshold is dependent on the aggregation strategy employed, with FedTrim exhibiting lower tolerance to malicious behavior than FedAvg and FedAvgM. Furthermore, the proposed framework converges faster than Krum while simultaneously ensuring transparency and decentralization through its design. The integration of consensus mechanism with neighborhood deviation provides a compelling alternative to existing solutions, 

\section{Conclusion and Future Work} \label{sec6}
This paper presented JiRAIYA, a Web3 enabled hierarchical FL framework designed to mitigate the impact of malicious participants while reducing reliance on enterprise grade centralized servers. By introducing Managers that orchestrate FL process within individual federations, JiRAIYA decentralizes coordination and improves scalability without sacrificing robustness. This proposed framework leverages the synergy of OVSVM based ND and Snowball consensus protocol to evaluate than accept model updates. Only consensus approved model updates are aggregated to generate the global model. In scenarios where no model updates satisfy consensus requirements, a reputation based fallback mechanism is employed. This mechanism, implemented using smart contracts, ensures transparency, immutability and fairness in determining model credibility. To preserve participant privacy, PCA based representations are used to encode model updates before evaluation, limiting exposure of sensitive information.

Extensive experiments conducted under malicious settings using various aggregation functions demonstrate the presence of adversarial participants degrades model performance, with notable impact observed when the Flip value is set to 4. Furthermore, results indicate the model poisoning attacks are significantly more detrimental than data poisoning attacks. Overall, JiRAIYA demonstrates that combining hierarchical coordination, blockchain based consensus and reputation mechanism can substantially enhance the robustness and deter malicious attack.

While JiRAIYA demonstrates promising results, several directions remain for future enhancement. One important extension is the introduction of adaptive aggregation thresholds, where a tunable hyperparameter dynamically determines the number of model updates considered during aggregation, similar to the Krum mechanism, enabling better resilience under varying levels of adversarial participation. Another key area of improvement lies in the development of a unified and standardized encoding strategy for model updates. The current reliance on PCA based representations may lead to inconsistencies if participants independently generate their own encodings potentially affecting fairness and comparability during evaluation. Moreover, ensuring the integrity of Manager reported outcomes is crucial. This can be achieved by incorporating digital signature based verification to guarantee that aggregation and consensus results are authentic and tamper resistant.





\clearpage 






\bibliographystyle{cas-model2-names}

\bibliography{swami}



\end{document}